# Comparing scalable strategies for generating numerical perspectives

HANCHENG CAO, Stanford University, United States
SOFIA ELENI SPATHARIOTI, Microsoft, United States
DANIEL G. GOLDSTEIN, Microsoft, United States
JAKE M. HOFMAN, Microsoft, United States

Numerical perspectives help people understand extreme and unfamiliar numbers (e.g., $330 billion is about $1,000 per person in the United States). While research shows perspectives to be helpful, generating them at scale is challenging both because it is difficult to identify what makes some analogies more helpful than others, and because what is most helpful can vary based on the context in which a given number appears. Here we present and compare three policies for large-scale perspective generation: a rule-based approach, a crowdsourced system, and a model that uses Wikipedia data and semantic similarity (via BERT embeddings) to generate context-specific perspectives. We find that the combination of these three approaches dominates any single method, with different approaches excelling in different settings and users displaying heterogeneous preferences across approaches. We conclude by discussing our deployment of perspectives in a widely-used online word processor.

CCS Concepts: • **Human-centered computing** → **Empirical studies in HCI**.

## 1 INTRODUCTION

How much is $330 billion? Like other extreme quantities (for example, a distance of 34 parsecs), unfamiliar dollar amounts can be hard to fathom without comparison to something else [7, 18]. To address this issue, it can be useful to employ *perspectives*: re-phrasings of measurements that make them easier to understand, via a change of units to express the *focal number* on a different scale, or a comparison to a *reference object*. For instance, $330 billion can be re-expressed using perspectives of "about $1,000 per person in the United States" or "about 5% of the United States Federal Budget". In addition to being intuitively appealing and perceived as helpful [6, 10, 13], perspectives have been shown to aid numerical comprehension by boosting recall, estimation, error detection, and prediction [3, 12, 27], which could find relevance in a wide variety of downstream applications.

These demonstrations of the benefits of perspectives have led to questions around what makes some analogies better than others, and if and how one can generate high-quality perspectives at scale for naturally occurring mentions of measurements. Approaches to automated perspective generation have varied, but they generally rely on first constructing a database of reference objects to compare measurements to and then prioritizing analogies to these reference objects that are both familiar and helpful to the reader [10, 27]. Prioritizing reference objects is complicated by the fact that what is most helpful for understanding a measurement can be difficult to quantify and can depend on the context in which the measurement occurs. For instance, if the $330 billion

Authors' addresses: Hancheng Cao, Computer Science, Stanford University, California, United States, hanchcao@stanford. edu; Sofia Eleni Spatharioti, Microsoft, United States, sspatharioti@microsoft.com; Daniel G. Goldstein, Microsoft, United States, dgg@microsoft.com; Jake M. Hofman, Microsoft, United States, jmh@microsoft.com.

discussed above were mentioned in an article about China's investment in renewable energy, it might be preferable to re-express the measurement as "about 13% of the Chinese government's yearly expenditures" instead of comparing it to U.S.-based references. Automatically generating such *contextual perspectives* is particularly challenging because it means having a database large enough to contain references that might apply across a wide range of contexts, but also having the precision to prioritize a relatively small subset of those references for any particular setting. To this point, while there is past work on automatic perspective generation [6, 10, 13, 27], in their current form none of these approaches generate such specific contextual analogies as those mentioned in the example above. As such, it is unclear how feasible it is to automatically generate high-quality contextual perspectives, and how the relative costs and benefits of doing so compare to simpler approaches that have been proposed in the literature.

In this paper we present and compare three policies for large-scale perspective generation in order of increasing complexity. We do so with an eye towards inclusive approaches that can be adapted to both other measurements, and to other countries and markets, and consider this when thinking about the generalizability of each policy. We restrict our attention to policies that can be deployed in real-time, interactive systems at scale with low latency and reasonable computational requirements at runtime. Based on a thorough review of the literature, our work appears to represent the first comparative analysis to weigh the costs and benefits of different approaches to perspective generation, and appears to be the first that develops a scalable method for contextual perspective generation that satisfies these requirements. In Section 3 we provide detailed information about each of the three policies we explore, but give a high-level overview here.

On one extreme we look at simple **rule-based policies** that use a fixed formula to re-express numbers. Rule-based policies differ depending on the type of measurement, but generally require at most one reference object for an entire domain, so are quick to construct and easy to scale. For instance, converting raw amounts to per-capita amounts is a rule-based policy. In the example above, one needs only the overall U.S. population to create a U.S. per-capita perspective for any dollar amount. Despite their simplicity, perspectives from rule-based policies have been effective in helping people understand matters from fuel efficiency to government expenditures [5, 9, 14]. Perspectives from rule-based approaches have also been shown to help people translate and rescale nonlinear attributes for better evaluability [11]. Nonetheless, these types of policies lack any explicit consideration of context, so it is an open question as to how often these perspectives are appropriate or helpful to readers when broadly deployed. Additionally, when used as a blanket policy, readers may eventually become tired of seeing the same type of perspectives and even learn to ignore them.

We then develop two, more sophisticated approaches for perspective generation that build on recent work. First, we consider a **crowdsourcing** policy inspired by Barrio et al. [3]. In this approach we construct a database of reference objects by directly asking one set of crowd workers to suggest familiar references for different quantities, and have another set of crowd workers rate them for familiarity and helpfulness, and a third set verify them. This policy has the potential to generate broad, common knowledge reference objects that readers are likely to be familiar with but that might not be present in conventional databases. A crowdsourcing approach requires substantially more time and effort to develop than a rule-based approach, but yields a greater number and variety of reference objects for constructing perspectives. At the same time, the policy is limited in that it is only cost effective for generating a modestly-sized database of reference objects, rendering this policy non-contextual in nature.

Finally we build on past work [6, 10] to develop a novel **contextual** approach to perspective generation that leverages open source knowledge bases along with large-scale language model embeddings. This policy, optimized for contextual relevance, is substantially more flexible than the rule-based and crowdsourced approaches, but also more complicated. It requires extracting and processing more than ten thousand reference objects from Wikipedia, and then learning a function to rank them for helpfulness and match them to the context in which a measurement appears. On one hand, the size of these resulting database holds promise, providing a wide range of possible reference objects. On the other hand, many of these objects are highly encyclopedic (e.g., the total expenditure of School District 29 in Dufur, Oregon), with no explicit signal that differentiates esoteric references that are unlikely to help readers from references that hold broader appeal. We show that web traffic data can be used as a scalable proxy for familiarity to enable a successful contextual policy.

To compare and evaluate our approaches, we focus on one type of measurement, value-based measurements expressed in dollar amounts appearing in news articles published in the U.S. We direct our attention to dollar amounts because they are the dominant form of measurements mentioned in the news media. For instance, looking at 20 years of published New York Times articles [29], there are roughly 3 million mentions of dollar amounts (with roughly 1 million of them containing values over $1 million) compared to fewer than half a million mentions of distances—which is the next most frequently mentioned unit—and fewer than 200,000 mentions of lengths, areas, and volumes combined. Dollar amounts are also critical for business, politics, and everyday decision making. We note that although we conduct our evaluation using value-based measurements, all three policies can be readily applied to other types of measurements.

In the remainder of the paper we first describe how each of these policies are constructed in detail and then present a preregistered study to compare them by asking participants to play the role of an editor who is trying to improve the readability of news articles. We also conduct a second, qualitative study to further identify motivations, preferences and perceived benefits for all policies. Somewhat surprisingly, we find that the simple rule-based approach is a competitive baseline, performing about as well as the contextual policy on average. And while both of these policies outperform the crowdsourced approach, interestingly the combination of the three strategies dominates any single method. We show that this is due to heterogeneous preferences for the type of perspectives preferred across different settings and by different users. We conclude by discussing how this work has informed our design and deployment of an end-to-end system for automatic perspective generation in a widely used online word processor.

## 2 BACKGROUND AND RELATED WORK

There is a rich literature demonstrating the importance of improving numerical comprehension among laypeople [2, 4, 23, 24], and many individual studies showing that simple, *rule-based policies* for presenting the same quantitative information in different formats can substantially impact comprehension and decision making in specific settings. For instance, Larrick and Soll [14] demonstrated that expressing fuel consumption as "gallons per 100 miles" instead of "miles per gallon" led to improved decisions around fuel efficiency. Dowray et al. demonstrated that re-expressing calorie amounts as number of miles needed to walk to burn them helped people better understand their calorie consumption [9]. Nguyen et al. [22] found that rounding numbers can greatly improve people's ability to remember numbers they have read and to make estimates based on them. Most recently, Boyce-Jacino et al. [5] showed that simple per-capita rescaling of large, government-related numbers improved numerical comprehension.

Here we emphasize that while each of these rule-based policies consists of a simple, fixed formula and applies to a specific type of measurement, each has been shown to be surprisingly effective in their respective domains.

Beyond the specific settings investigated in these studies, there have been several attempts to study the benefits of perspectives more generally, and to develop systems for automatic perspective generation, nearly all of which has come out of the human-computer interaction community, as described below.

Barrio et al. [3] served as a proof-of-concept for the benefits of perspectives in general. The authors developed a templating system for re-expressing numerical measurements using a *crowdsourced* policy, and showed that crowd participants were capable of using these templates to manually generate high-quality perspectives for individual news articles. They tested a handful of such perspectives in randomized experiments, demonstrating that perspectives have substantial benefits for people's ability to estimate, recall, and detect errors in published numbers. This work did not, however, explore methods for automatically generating perspectives at scale.

Kim et al. [13] was one of the first attempts to generate automatic perspectives, focusing on analogies for spatial measurements (areas and distances) mentioned in text articles. The approach here was around personalization, suggesting landmarks and geographic areas that were likely to be familiar to the user based on their location (e.g., rephrasing an area of 50 square kilometers as about four times the area of Golden Gate park for someone located in San Francisco). The authors used a hand-tuned ranking function to prioritize analogies, and found that users rated personalized analogies from their system higher than generic ones. Many of the terms in this ranking function were specific to spatial analogies (e.g., distance to a landmark or familiarity as measured by Flickr photo counts), and not easily adapted to other dimensions, including dollar amounts. Additionally this work did not consider the context in which a spatial measurement was mentioned when generating analogies. That is, the same analogy would be used for all news stories conditional on the measurement and the user's location.

Riederer et al. [27] developed a systematic approach to generating perspectives, with a focus on instant answer facts found in search engines, specifically the areas and populations of different countries (e.g. "100 million acres is about the area of California"). Here the authors used a series of relatively expensive crowdsourced estimation tasks to gauge people's familiarity with a clearly defined and relatively compact class of reference objects (U.S. states), and fit a domain-specific ranking model. They showed that perspectives from this system had long-lasting benefits on comprehension of up to six weeks, but as with work by Kim, et al., the approach does not readily generalize to other dimensions such as dollar amounts, especially when there is an ill-defined reference class, and does not consider context when generating perspectives.

Hullman et al. [10] presented a method for automatic perspective generation for physical measurements (weight, height, length, and volume) that leveraged several distinct data sources (WordNet, ImageNet, Amazon products, and DBPedia/Freebase) to construct a set of familiar reference objects, and used a hand-tuned ranking function to prioritize analogies (e.g., 28 lbs is about the weight of a microwave). They conducted experiments to show that these analogies improved reader comprehension and that readers generally found perspectives helpful. This approach is closer to ours in spirit, but not easily adapted to other domains and does not generate contextual perspectives.

*Contextualized* policies are explored in Chaganty and Liang [6] focusing on the natural language processing aspect of perspective generation. Specifically, the authors developed a recursive neural network to generate "compositional perspectives" that translate multiplicative combinations of different reference amounts into a natural-sounding perspective phrase. They used a small database of fewer than 200 reference objects, but their ability to combine reference objects allows the system to cover a wide range of measurements. The authors showed that the statements generated by this system are more natural-sounding (have a higher BLEU score) than those of a simpler baseline for translating multiplicative formulas to text. The authors also conducted a small scale study of user preferences for these compositional perspectives, finding that the most common cause of errors noted by users is a mismatch between the perspective and the context in which the number appears. They attribute this to the small size of the knowledge base that they constructed. Our approach is to use the key insights from this work to generate contextually relevant perspectives by focusing on simpler (non-compositional) perspectives, with a substantially (roughly 30 times) larger reference database, inspired by [10]. This approach avoids the computational overhead of the recursive neural network approach, but increases the need for good signals for familiarity, which we discuss below.

| Paper | Approach | | | Ranking process | Contextual | Generalizes to other dimensions |
|---|---|---|---|---|---|---|
| | *Rule-based* | *Crowdsourced* | *Database* | | | |
| Larrick & Soll, Science 2008 | ✓ | | | N/A | | |
| Dowray et al., Appetite 2013 | ✓ | | | N/A | | |
| Boyce-Jacino, PNAS 2021 | ✓ | | | N/A | | |
| Barrio et al., CHI 2016 | | ✓ | | N/A | | ✓ |
| Kim et al., CHI 2016 | | | ✓ | Hand-tuned | | |
| Riederer et al., CHI 2018 | | | ✓ | Learned | | |
| Hullman et al., CHI 2018 | | | ✓ | Hand-tuned | | |
| Chaganty & Liang, ACL 2016 | | | ✓ | Learned | ✓ | ✓ |
| This paper | ✓ | ✓ | ✓ | Learned | ✓ | ✓ |

Table 1. Summary of related work on methods for generating numerical perspectives.

Table 1 provides a summary of the prior research on perspectives discussed above. As the bottom row indicates, our work is the first to compare all three approaches for perspective generation to each other directly. Notably, this includes a comparison to simple rule-based policies, which have previously been omitted by work in the computer science community. In addition, our work uses a learned—rather than hand-tuned—ranking function to select the best perspective for a given measurement, and takes into account the context provided by the words surrounding a measurement. Finally, while we focus on dollar amounts here, the approach we present can be easily generalized to other dimensions, unlike much past work on perspectives. We discuss generalizability further in Section 5.3.

## 3 POLICIES FOR GENERATING PERSPECTIVES

In this section, we describe the three different policies we developed and compared for generating perspectives of dollar amounts: a rule-based approach, a crowdsourced system, and a contextualized perspective approach. As a preview, Table 2 shows the perspectives generated by each of the three policies for three different sentences, all containing the same focal measurement of $100 million. Note that the rule-based and crowdsourced perspectives are identical for all three quotes because they depend only on the focal measurement, whereas the contextual perspectives change based on the words surrounding the measurement.

### 3.1 Rule-based perspectives

We first implement a simple rule-based policy to generate perspectives for large dollar amounts: per-capita normalization relative to the U.S. population. Past work has shown that rescaling large numbers in this way is effective at improving numerical comprehension in particular settings [5], but to the best of our knowledge this approach has not been tested as a generic method for perspective generation. The policy relies on only one reference number (325 million people) and simple arithmetic. Given some dollar amount, we divide this number by the total population size, and then round the result to two significant digits, following recent work on the cognitive benefits of round numbers over precise ones [22].

| Quote | Rule-based (Per-capita) | Crowdsourced | Contextual |
|---|---|---|---|
| The U.S. cut its military budget by $100 million. | about $0.30 per person in the United States | about the cost of a private high-end jet | about 0.01% of the United States military budget |
| The Cinncinati Reds spent over $100 million on salaries this year. | about $0.30 per person in the United States | about the cost of a private high-end jet | about 4% of the total U.S. baseball salaries for all teams |
| Walmart acquired the merchandising startup for $100 million. | about $0.30 per person in the United States | about the cost of a private high-end jet | about 0.02% of the annual revenue of Walmart |

Table 2. Example perspectives generated by each of the three policies we implement and compare. Rule-based and crowdsourced perspectives depend only on the focal measurement, whereas contextual perspectives adapt to the surrounding text.

## 3.2 Crowdsourced perspectives

The second policy we explore is a crowdsourced approach to perspective generation. The crowdsourced perspectives used in this work were the result of a broader effort in generating a crowdsourced set of items to cover a wide range of measurements across multiple dimensions. To generate this set, we created a standalone crowdsourcing platform, with three main components: a) a proposing objects component, b) a rating objects component and c) a verification component. Combined, the three components make up a system where the crowd can collaborate on surfacing high quality perspectives for a wide range of measurements. A diagram of the platform can be found in Fig. 1. Although our system has been readily ported to other measurements, such as area, height, etc., in this section we focus on the process of generating crowdsourced perspectives for dollar amounts only.

*3.2.1 Generating reference objects .* To generate an initial crowdsourced list of candidate objects that could potentially be used as perspectives, we utilized the *proposing* objects component to conduct a study through Amazon Mechanical Turk. We invited participants to come up with objects that could be helpful in describing a given measurement as a crowdsourced step of ideation [28]. In addition to the object, we also asked participants to provide a reference page that could support their claim, as well as the option to link their suggested object to a Wikidata[1] entity using the Wikidata Search API[2]. Screenshots for the proposing component can be found in Fig. 2A.

Participants were asked to propose objects for up to 20 randomly drawn measurements, and were given the option to quit at any time to receive payment. Measurements were randomly drawn from a pool of 25 possible dollar amounts options, spanning 13 orders of magnitude from $1 to $1 trillion on a rounded logarithmic scale (e.g., $1, $3, $10, $30, ...). We used a quality-based payment scheme to incentivize high quality suggestions. More specifically, participants received a base payment of $0.25 for completing the study, plus a bonus per object suggested. The bonus was tied to the average helpfulness score the object would subsequently receive via the rating component, with

---

[1]Wikidata is a popular knowledge base widely used in both research and applications[31]. https://www.wikidata.org/wiki/Wikidata:Main_Page

[2]https://www.wikidata.org/w/api.php?action=wbsearchentities&search=

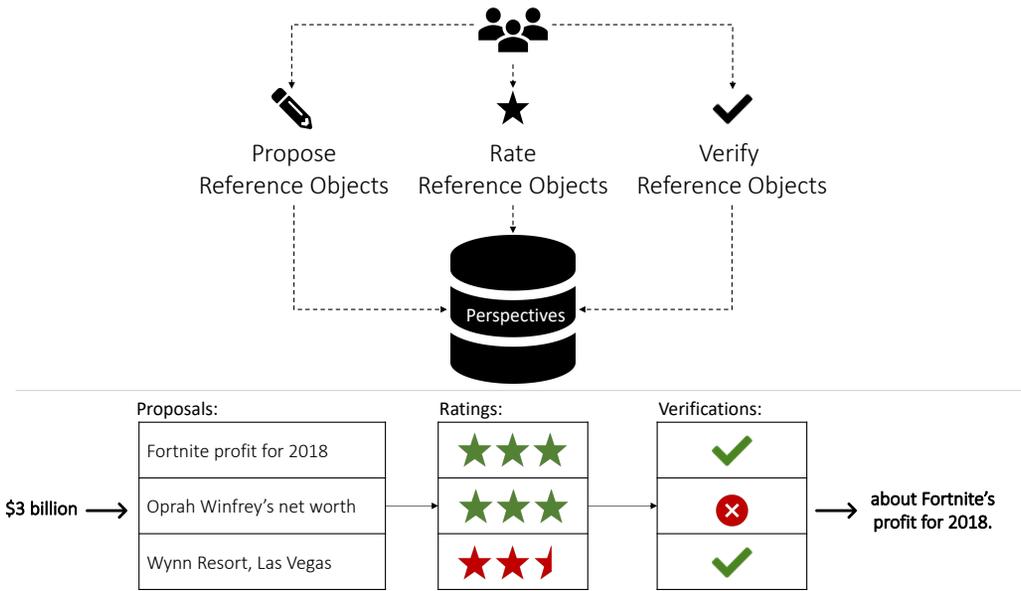

Fig. 1. Overview of the crowdsourced platform for generating perspectives. Candidates proposed are independently rated for quality and verified for accuracy, resulting in the final database of reference objects.

five different bonus levels, $0, $.05, $.10, $.25 and $.50 corresponding to helpfulness scores of 1 through 5. We collected 145 candidate reference objects for dollar amounts from 118 participants overall. For HIT qualifications, we required U.S. based participants with a 95% approval rate.

*3.2.2 Rating reference objects.* Having collected our set of candidates, we then proceeded to collect *ratings*, in order to identify the highest ranked objects. To do so, we invited participants who had not previously participated in the proposal stage to rate candidate objects using the rating component of the crowdsourcing platform. The exclusion requirement was necessary in order to prevent participants from providing favorable ratings to their own proposals, therefore achieving higher bonuses. We asked participants three questions. First, we showed them a quote comparing a measurement to a candidate object, and asked them to rate on a scale of 1 through 5 how helpful they felt the comparison was for understanding the measurement. We refer to this rating as the *helpfulness rating*. Next, we asked participants to rate on a scale of 1 through 5 their familiarity with the specific dimension of the candidate object. We refer to this rating as the *familiarity rating*. Screenshots for the rating component can be found in Fig. 2B.

Participants were asked to rate up to 20 different, randomly drawn, candidate objects, and were given the option to quit at any time to receive payment. We used a fixed bonus payment scheme, consisting of a base fee of $.10 for completing the ratings study, plus a fixed bonus of $.05 for each object rating provided. Again, we required U.S. based participants with a 95% approval rate. We received ratings for dollar amounts from 247 participants. We then compiled a total rating for each object, by combining the average helpfulness and familiar ratings. Objects that received at least 5 ratings and had a total rating of at least 3 were deemed "acceptable". This process resulted in 71 acceptable reference objects for dollar amounts.

*3.2.3 Finalizing perspectives.* Recall that in the proposal stage, we required participants to attach a reference page connecting their proposed object to a given measurement. This measure was designed to incentivize *accurate* objects, such that the size of the object was within a reasonable distance to the measurement it was proposed for. However, it is still possible for some items in our acceptable objects subset to not be accurate, either due to the proposed object being way off the measurement, or due to high variance. Consider the case where someone suggests a car for a given value. While this object may receive high ratings, cars can vary heavily depending on their type, maker etc. A compact sedan may range from $15,000 to $30,000, while some luxury cars can be valued above $100,000. Therefore, an additional step is required to ensure that our perspectives are both highly rated and accurate.

Fig. 2.   Crowdsourced Perspectives Interface. (A) In the proposal stage, participants are asked to suggest objects for given measurements and optionally link suggestions to Wikidata entities. (B) In the rating stage, proposals are rated for helpfulness and familiarity. Items with a combined rating above 3 are deemed acceptable. (C) Proposals are also independently verified for accuracy.

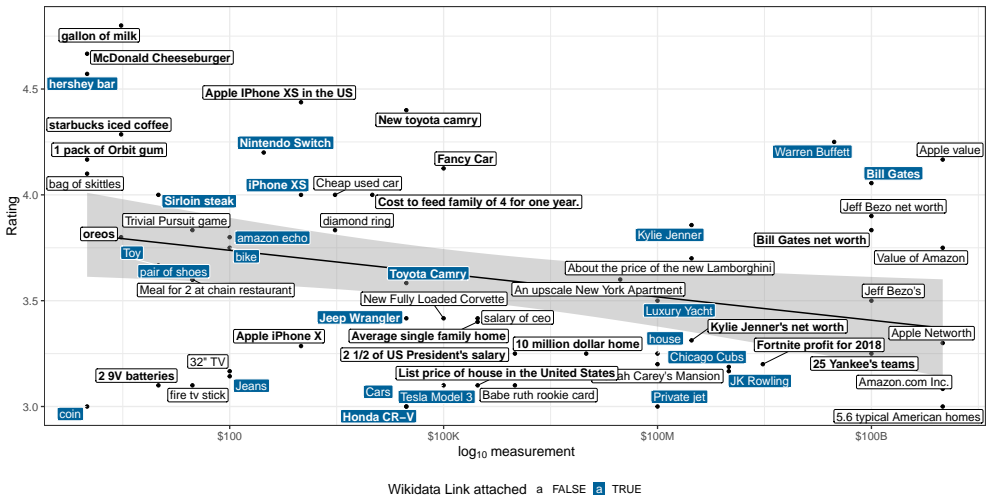

Fig. 3. Dollar amounts perspectives generated from our crowdsourcing system, with a total rating above 3. Bold indicates the object was successfully verified by participants. Objects highlighted in blue had a Wikidata entity attached by participants during the proposal stage.

To evaluate the accuracy of the top-rated reference objects, we invited participants, to provide *verifications*, again excluding those that have completed any of the previous studies. We first asked participants to estimate the size of a given object without looking it up. We then provided participants a search engine for them to use, with the goal of finding a website that contains the true size of the object. Finally, we asked them to input the value they found. Screenshots for the verifying component can be found in Fig. 2C.

We collected verifications from 202 participants through Amazon Mechanical Turk, using a similar payment setup as the ratings study, with a base pay of $0.25 for completing the study, $0.2 bonus per submitted verification, and ability to quit at any time to receive payment. We used the same qualifications as before, requiring U.S. based participants with a 95% approval rate. An object was deemed "accurate" if it had received at least 3 verifications and its median verified amount was within 20% of the measurement it was suggested for. The verification stage yielded 34 reference objects for dollar amounts, that were both acceptable and accurate.

We observed that proposals for dollar amounts were less likely to be additionally associated with Wikidata entries, with only 30% of the proposals having a Wikidata link made by users at this stage. Examples of suggestions that could not be tied to a specific Wikidata entry at all included 'cheap used car', 'salary of a CEO' and 'cost to feed a family of 4 for one year'. In addition, we also received Wikidata-linked suggestions for general category items, for which no definitive measurement may be present in Wikidata records, such as 'bicycle helmet', '32" TV' and 'luxury yacht'. Suggestions belonging to both of these cases indicate that our crowdsourcing approach can surface common-sense suggestions not present, or clearly defined, in large scale knowledge bases. Fig. 3 provides an overview of acceptable reference objects for dollar amounts, including information on crowd verified accuracy and Wikidata connectedness. The final subset was manually reviewed by the authors for duplicates, phrase cleanup, and crowd verification final determinations.

### 3.3 Contextual perspectives

The third and final policy we implemented is a contextual one that generates perspectives based on an input measurement and the sentence in which that measurement appears. At a high level, developing this policy involves four steps: 1) compiling a database of candidate reference objects that can potentially be used as perspectives, 2) specifying features for a helpfulness model to rank reference objects given an input sentence, 3) collecting training data from crowd participants on how helpful they find different perspectives, and 4) fitting the helpfulness model to the crowdsourced training data. In short, we extract reference objects from two freely available sources: the Dictionary of Numbers [8] and Wikidata [31]. Then we specify a ranking model that considers the semantic similarity between an input sentence and each reference object, the familiarity of each reference object, the type of each reference object, and the multiplier used in the perspective. We use Sentence-BERT [25] to capture semantic similarity and Wikipedia traffic data to impute a proxy for familiarity. We devise a data collection task where we ask participants to rate the helpfulness of various candidate perspectives for a stratified sample of news articles spanning different topics, and containing numbers of different magnitudes. Finally, we use the results of this crowdsourced task to train the ranking model, which provides the final contextual perspective policy that we evaluate in the next section. A visualization for the process of generating contextualized perspectives is shown in Fig. 4. We discuss each of these steps in more detail below.

*3.3.1 Generating candidate reference objects.* Reference objects were collected from two sources. First was the Dictionary of Numbers, a browser extension that detects numbers in Web pages and inserts inline perspectives next to them to help readers. The package contains a small database of hand-selected reference objects collected from a variety of sources. We extracted all 253 dollar amounts in the database as candidate reference objects.

Second, we collected reference objects from Wikidata, the same open source knowledge base used in developing the crowdsourced policy. In Wikidata, measurements for entities are stored as *properties* (such as width or volume), with units (such as meters or liters). We selected Wikidata entities with properties expressed in the *currency* unit as of July 2020[3]. We pre-processed the reference objects as follows. First, we removed non U.S. dollar reference objects, for example entities denominated in Euros. Second, we removed reference objects with negative currency values. Third, for entities that were measured at multiple times (e.g. yearly U.S. GDP values), we took only the most recent value. Fourth, we manually checked different properties of categories and removed references from obscure or redundant categories. For example, we chose 'nominal GDP' over 'GDP (PPP)', and removed references to 'total liabilities.' Applying these pre-processing measures resulted in a final dataset of 3,975 Wikidata-sourced reference objects.

We then combined reference lists from the Dictionary of Numbers and Wikidata sources to make a base set of 4,228 candidate reference objects to use in contextualized perspectives, as summarized in Table 3.

*3.3.2 Specifying a ranking model.* After collecting a list of candidate reference objects for generating contextualized perspectives, we developed a regression model that takes as input the focal number $n$ (which is the numerical value of the measurement, e.g. 3.7 billion in Fig. 4), the text (that is, context) surrounding the focal number, and a candidate reference object with its numeric value $n_{ref}$. It outputs a predicted, context-specific 'helpfulness score' for each candidate reference object. The model can then be used to select a reference object $r$ with the highest predicted helpfulness score, and to re-express the focal number as *multiplier* times the value of $r$, where *multiplier* = $\frac{n}{n_{ref}}$. We round the multiplier to one significant figure, and format the final perspective as:

---

| Source | Property | Count | Examples |
|--------|----------|-------|----------|
| Dictionary of Numbers | Dictionary | 253 | the cost of average used car, the typical CEO pay |
| Wikidata | Nominal GDP | 223 | the nominal GDP of the United States |
| | Nominal GDP per capita | 206 | the nominal GDP per capita of the United States |
| | Annual budget | 64 | the annual budget of Boston Police Department |
| | Cost | 46 | the cost of World Trade Center |
| | Endowment | 22 | the endowment of Harvard University |
| | Market capitalization | 71 | the market capitalization of Amazon |
| | Net profit | 152 | the net profit of Apple Inc. |
| | Price | 55 | the price of Manhattan Center |
| | Total assets | 144 | the total assets of Verizon |
| | Annual revenue | 2931 | the annual revenue of Metropolitan Museum of Art |
| | Total equity | 61 | the total equity of Delta Air Lines |

Table 3. Reference objects for generating contextual perspectives by property.

- 'about the same size as the value of $r$' if *multiplier* is rounded to 1,
- 'about X times the value of $r$' if *multiplier* is rounded to be greater than 1, and
- 'about X% the value of $r$' if *multiplier* is rounded to smaller than 1.

We chose the following functional form to capture semantic similarity between the input text and each reference, the familiarity of the reference, the type of reference, and the multiplier:

$$\begin{aligned} \text{helpfulness} \sim & \beta_s \text{similarity(reference,sentence)} + \\ & \beta_f \text{familiarity(reference)} + \\ & \beta_{\text{category(reference)}} + f(\text{multiplier}) \end{aligned} \quad (1)$$

The $\beta$ coefficients represent the relative weights given to each of these terms, which we learn from crowdsourced training data as described in the following section. Below we describe how each of these four features are computed.

*Contextual similarity.* To quantify similarity between reference objects and the input sentence, we adopt a Sentence-BERT embedding approach [25][4]. SBERT belongs to the family of Transformer-based models recently proposed and widely used in natural language processing [32]. Sentence-BERT transforms sentences into vector embeddings. Here we leveraged the pre-trained 'paraphrase-mpnet-base-v2' model, which was trained using the paraphrase dataset and has been shown to be an effective model for sentence embedding [26]. We generated sentence embeddings for each of the candidate reference objects in our database. Then, for a given sentence containing a focal number, we compute the embedding for that sentence and take the cosine similarity between its embedding and each reference object. This comprises the similarity term in the ranking model. We also compared the Sentence-BERT approach with more classical methods (e.g., pre-trained word2vec [20], and edit distance) to measure textual similarity through retrieving the most relevant reference object for a given quote. We found that Sentence-BERT clearly outperformed all other methods in terms of the quality of retrieved objects.

*Familiarity.* To quantify the familiarity of a reference object, we used Wikipedia pageviews as a proxy of familiarity: the more monthly pageviews an entity has on Wikipedia, the more likely it is to be familiar to readers. Wikidata entities often provide a link to their corresponding Wikipedia pages. We used the Wikipedia API[5] to retrieve pageviews in Jan 2021 for entities appearing in our candidate reference object lists, and successfully retrieved pageviews for 71% of the reference objects. Well-known entities such as 'Youtube', and 'Google' ranked towards the top, while lesser-known entities such as 'Bishop Museum' and 'Aspen Music Festival and School' ranked lower down. To impute a familiarity score for reference objects without Wikipedia pageviews (that is, reference objects coming from the Dictionary of Numbers and Wikidata entities with non-existent or broken Wikipedia links), we fit a ridge regression model to predict Wikipedia page views (logged) from reference object BERT embeddings as contextualized word embeddings are known to implicitly encode word frequency information through the training process [16, 33]. On a held-out test set the model's $R^2$ was 0.48 on a test set, indicating good potential for predicting familiarity from reference object embeddings. For consistency, we used the ridge regression model to predict the log page view of each reference object, and use this predicted value as a proxy of familiarity in the ranking model.

---

[4]Implementation and pretrained model is available at: https://www.sbert.net/
[5]Wikimedia Foundation Pageview Tool: https://github.com/mediawiki-utilities/python-mwviews

*Category.* The property category in Table 3 is likely to influence the perceived helpfulness of perspectives as well, with some categories being more helpful than others. Certain property categories may be more familiar to people than others. We thus consider the property as a categorical variable in the ranking model. For reference objects coming from dictionary sources, and thus had no property name attached, we created a 'Dictionary' value.

*Multiplier.* The value of the multiplier could also play a role in helpfulness. Intuitively readers may prefer multipliers that are neither too large nor too small [5, 13], perhaps because they facilitate estimation and reasoning [13, 27]. We entered the log transformed multiplier and its squared term (first and second order polynomial terms of the log transformed multiplier) into the helpfulness score rating model to capture potentially non-linear preferences over multiplier values.

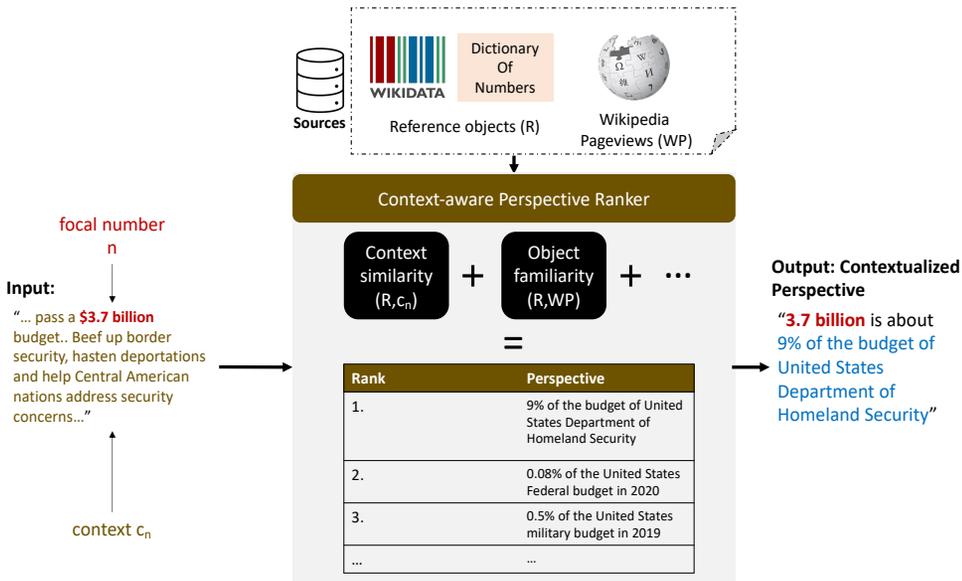

Fig. 4. Process for generating contextualized perspectives. A context-sensitive regression model scores reference objects by predicted helpfulness. Helpfulness is predicted based on context similarity, object familiarity, type of reference and multiplier. The process outputs the predicted most helpful option in the form of a perspective.

*3.3.3 Collecting training data.* To train and evaluate our perspective ranking model, we first needed to collect "ground truth" data on people's judgments about how helpful candidates perspectives are in various contexts. To this end, we recruited participants through Amazon Mechanical Turk to complete a task in which they were presented with quotes in context and asked to rate a list of candidate perspectives according to how helpful they are for understanding the numerical values.

*Stimuli.* For the stimuli in this task, we extracted quotes containing dollar amounts from articles that appeared on the front page of the U.S. edition of the New York Times in 2017. A large number of quotes was necessary in order to have enough semantic context to cover various categories. The quotes were sourced from the following eight sections: Art, Business, Health, Science, Sports, Technology, U.S. and World. The magnitude of the dollar amounts in these quotes ranged from $10^6$ (1 million) to $10^{13}$ (10 trillion), representing approximately one third of all dollar amounts published in the newspaper. In total, we collected 5,128 quotes.

To assess helpfulness, participants would ideally be presented with a quote and asked to rate the helpfulness of perspectives generated from the set of *all* possible reference objects. However, such approach is not feasible in practice as there are over 4,000 candidate reference objects available for each quote and it is not practical to ask participants to read through and rate all of them. Instead, we provided people with the top 10 reference objects based on closest semantic relevance to each quote context (determined using the Sentence-BERT method mentioned in Section 3.3.2). This allowed us to collect feedback on candidate perspectives that were reasonable in nature and would be informative in training the ranking model.

We then placed quotes into 64 different bins, covering different topics (8 sections of the news paper) and magnitudes of focal numbers (8 powers of ten, from millions to 10 trillions). From each bin we took up to 3 quotes whose top ranked reference object had the highest similarity score to the quote. (Note that sometimes such quotes may not exist, as certain sections may not have very small or large magnitude dollar amounts.) We further cleaned the selected quotes (e.g., removing quotes that were too short or too vague to provide enough context, and removing wrongly parsed phrases in quotes such as 'all rights reserved.'), resulting in 180 quotes in total. We then randomized these quotes into 10 equal sized bins (each with 18 quotes), and generated 10 contextualized perspectives based on the top 10 semantically closest reference objects to the selected quotes. We provide more details on quote selection in Appendix A.

*Procedure.* The interface used in the data collection task is shown in Fig. 5. Participants were provided with a randomly selected bin of 18 quotes and corresponding perspectives for evaluation. We further randomized the order of the quotes shown to individuals to control for order effects. At the end of the evaluation task, participants were able provide feedback using an optional text entry field. For each item in the randomly assigned bin, participants were shown the full quote, and asked to answer two questions regarding the focal number contained in it. First, a yes/no question about whether they thought readers would benefit from seeing a perspective for this quote. Second, a list-based of at most 12 perspectives in random order: 10 contextualized perspectives generated by selecting top 10 semantically closest reference objects, 1 rule-based perspective if applicable, and 1 crowdsourced perspective if applicable. For each perspective, participants were asked to provide a helpfulness score, on a scale from 1 to 3, where 1 indicates that the perspective is neither relevant nor helpful, 2 means the perspective is somewhat relevant and helpful, and 3 means the perspective is very relevant and helpful.

In addition to these quotes, each participant saw one additional quote that served as an attention check. This attention check quote, which was the same for all participants, read "Congress was tasked with cutting $330 million from the United States Budget" and all but one of the accompanying

perspectives were designed to be non-sensical: "about 70% of the age of the universe", "about 70% of the speed of the B-2 Stealth aircraft", or "about 10 times the radius of the moon". The correct answer for this quote was to rate "about $1 per person in the U.S." more highly than the nonsensical distractors. The order of the attention check question was also randomized, and we removed training data from anyone who failed this check.

*Participants.* We compensated participants $4 for completing the study. Payment was determined based on our findings in pilot rounds that it typically takes participants less than 20 minutes to complete the task. We recruited participants through Amazon Mechanical Turk and restricted recruitment to those with a masters qualification to ensure high-quality responses.

Fig. 5. Screenshot of the interface used for collecting crowd evaluations. Participants were shown a quote with the focal number highlighted, then asked to rate a list of reference objects for helpfulness.

*Results.* In total, we recruited 295 participants; 243 (82%) of them passed the attention check. First, as shown in Fig. 6A, participants reported that perspectives would be helpful across all sections of the newspaper. We observed that as the focal number increased, so did the percentage of people who thought the number needed a perspective. Second, as shown in Fig. 6B, we found a positive relationship in general between semantic relevance of perspectives (cosine similarity between quote embedding and reference object embedding) and perceived helpfulness, meaning that more semantically relevant reference objects were more likely to be found helpful by people. The observed correlation was statistically significant but relatively weak, indicating that there were benefits to be had from the other terms in the ranking model such as familiarity and the category of the reference object. See Table 5 of Section B of the Appendix for more details.

### 3.3.4 *Fitting the ranking model.* After collecting training data using the task described above, we used it to fit the weights in the ranking model in Eqn. 1. Specifically, we fit a clipped linear

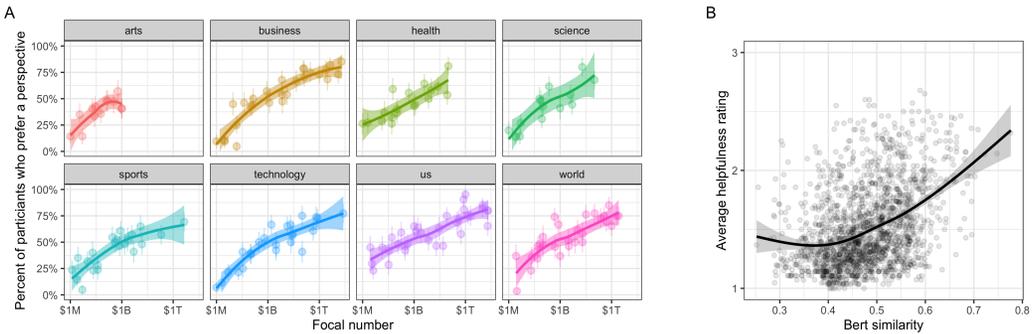

Fig. 6. (A) Preference for perspectives by magnitude of the focal number in each quote and the section in which it appeared. Points show the magnitude of the focal number for one quote along with the percent of participants who preferred a perspective to rephrase the focal number. Lines show the estimated average trend in each section, with bands showing a 95% confidence interval. (B) Average helpfulness ratings for each candidate perspective as a function of the semantic similarity between quotes and perspectives. Points represent perspectives, and the line shows the estimated average trend with a 95% confidence interval. All trends are estimated via LOESS regression.

regression using the Vowpal Wabbit package [6], which nicely handles the fact that we have a continuous helpfulness score on a bounded scale (from 1 to 3).

We examined four different versions of the model to assess the value of different terms in the model: 1) textual similarity only, 2) textual similarity + familiarity of reference objects + reference objects, 3) textual similarity + familiarity of reference objects + reference objects + log transformed multiplier, and 4) a textual similarity + familiarity of reference objects + reference objects + log transformed multiplier + squared of log transformed multiplier. The last model was designed to capture the idea that there might be a small range of particularly helpful multipliers, with worse multipliers lying both above and below that range.

All models took 50 passes to be trained. To test the models, we randomly divided our labelled data into an 80% training set and 20% test set, trained the model on the training set and compared prediction performance for the four different models. The $R^2$ for the four different models were 0.029, 0.104, 0.104, 0.118 respectively, indicating a considerable performance gain when adding familiarity and types of reference objects as predictors. However the gain from adding the multiplier was minimal, and manually inspecting results from the multiplier models showed top-ranked perspectives that were inferior to the other models.

Based on this evaluation, we chose the third model (with textual relevance, familiarity, and the property category) as the final model. Overall, we found that textual relevance plays an important role in increasing the helpfulness of a perspective: the higher the semantic relevance, the more helpful the perspective. The familiarity of the reference objects also plays a role. As expected, people found more familiar reference objects to be more helpful. The effect of different types of reference objects varied, with dictionary reference objects being more helpful, and the property categories of cost, endowment, and price less helpful.

---

[6]https://vowpalwabbit.org/

## 4  EVALUATING POLICIES

Having described three policies for perspective generation above, here we directly compare them in a crowdsourced study designed to simulate how a tool for automatic perspective generation would be used in authoring or editing content. In contrast to the task discussed in the previous section used to collect training data for the contextual policy, the goal here was to see how often participants would actually insert suggested perspectives from each of the three policies into text they were authoring. Participants were shown a series of quotes, each containing one measurement along with three perspectives (one from each policy) that could be added to quotes to help readers. For each quote they could either select one of these three perspectives or choose not to add any of them. We selected the quotes to be representative in that they span different topics (sections of a newspaper) and contain numbers of different magnitudes (from millions to trillions). Our key dependent variable is the *keep rate* for each policy (i.e., the number of times a perspective from a given policy is selected divided by the number of times it is shown).

Before running this study, we pre-registered the following analyses[7]:

- Comparing the overall keep rate for contextual vs. crowdsourced perspectives.
- Comparing the overall keep rate for contextual vs. rule-based perspectives.
- Analyzing how preferences for each perspective policy varies with the magnitude of the focal number being put into perspective.

We report all sample sizes, conditions, data exclusions, and measures for the analyses that were described in our pre-registration documents. We determined the sample size for this task in order to detect differences in keep rates as small as 4 percentage points in keep rate (observed in a pilot study) at 95% power with a 5% significance threshold, accounting for a loss of subjects due to our pre-registered exclusion critera.

### 4.1  Task Design

*4.1.1  Stimuli.* We again leveraged quotes containing dollar amounts that were published in the 2017 U.S. version of the New York Times. We first excluded all quotes that were used in prior rounds of data collection and model training. Then we performed a stratified random sample of quotes spanning eight sections of the newspaper and seven orders of magnitude (from millions to trillions, in powers of ten) to cover a broad range of contexts and focal numbers. We manually reviewed the sampled quotes and removed low-quality quotes (e.g., short sentences with little information other than a dollar amount), to arrive at a total of 136 quotes. We generated three candidate perspectives for each quote, one from each of the three policies we considered (rule-based, crowdsourced, and contextual). We randomly divided these quotes and their accompanying perspectives into 8 batches of 17 quotes each. We provide more details on quote selection in Appendix A.

In addition to these quotes, each participant saw one additional quote that served as an attention check. This attention check, which was the same for all participants, read "Congress was tasked with cutting $330 million from the United States Budget" and the accompanying perspectives were designed to be non-sensical: "about 70% of the age of the universe", "about 70% of the speed of the B-2 Stealth aircraft", or "about 10 times the radius of the moon". The correct answer for this quote was to prefer none of these perspectives, and we excluded any participants who answered otherwise.

*4.1.2  Procedure.* Each participant was assigned to see all of the quotes from one of the 8 batches described above, selected at random, plus the attention check quote, for a total of 18 quotes. Each quote contained one focal number along with three perspectives (one from each policy) that could

---

## Task 1 of 18

Imagine you are writing an article with the following sentence in it and are trying to help your readers best understand the number in the sentence.

*The Congressional Budget Office said in August that if the cost-sharing subsidies were cut off, premiums would shoot up 20 percent next year, and federal budget deficits would increase by $194 billion in the coming decade.*

Which of the following, if any, would you add to this article to help readers understand the number $194 billion?

| Perspective | Selection |
|---|---|
| about 4% of the United States Federal budget in 2020 | ○ |
| about $600 per person in the US | ○ |
| about 2 times the value of the net worth of Bill Gates | ○ |
| I would not add any of these perspectives to the article. | ○ |

Next

Fig. 7. Screenshot of the interface used for evaluating and comparing policies. Participants were shown a quote with the focal number highlighted and three perspectives (one from each policy considered). They were then asked which perspective, if any, they would add to help readers better understand the focal number.

be added to the quote to help readers. As depicted in Fig. 7, participants were told to imagine that they were writing an article containing each quote and asked which of the displayed perspectives, if any, they would add to the article to help readers better understand the focal number. For each quote they could either select one of these three perspectives or choose not to add any of them.

Quotes within each batch were shown in a randomly selected order. Within each quote the order of the three perspective options was randomized, with the option to prefer no perspective always shown last. Participants were not shown which perspective came from which policy.

*4.1.3   Participants.* We recruited participants who had not taken part in our previous data collection task, who are identified as Masters by the platform, and who had at least a 99% approval rating for having completed at least 100 previous tasks. We compensated participants $2.00 for completing the task, which took a median of 6 minutes. We recruited 70 participants in total and 62 (88%) of them passed the attention check.

### 4.2   Results

*4.2.1   Comparing each policy.* Overall we find that 35% of the time participants choose not to add a perspective, and the remaining 65% of the time they select one of the three perspectives, with keep rates of 24% for rule-based perspectives, 17% for crowdsourced perspectives, and 24% for contextual perspectives. Two-sided proportions tests show a statistically significant difference between the keep rate for contextual and crowdsourced perspectives ($\chi^2(1, n = 1054) = 18.11$, $p < .001$), but we found no evidence for different keep rates between contextual and rule-based perspectives ($\chi^2(1, n = 1054) = 0.09$, $p = .760$).

*4.2.2   The combination of three proposed perspective approaches dominates any single method.* We also examined how keep rates changed as we considered combining policies, starting with only the simplest policy first (rule-based), then adding crowdsourced perspectives, and finally presenting perspectives from all three approaches. This analysis is designed to assess the return on the investment to developing and combining increasingly sophisticated methods for perspective generation. As shown in Fig. 8, there are substantial returns to combining policies, and using all three strongly outperforms any single method. Specifically, looking at Fig. 8A, if we had only a rule-based policy, perspectives would be added roughly a quarter of the time (red), but displaying perspectives from all three policies leads to perspectives being chosen more than two thirds of the time (green). The larger points in this figure show the overall average keep rates under different policy combinations, whereas the smaller points show each individual participants' average keep rate across all of the quotes they would see in these settings. Fig. 8B provides another view of these per-participant keep rates, displaying the cumulative fraction of of participants who would be at or above a given average keep rate under different policy combinations. For instance, looking at the 50% mark on the x-axis shows that if we used only the rule-based policy, roughly 10% of participants would select a perspective at least 50% of the time (red), whereas by using all three policies, more than 80% of participants would have an average keep rate of at least 50% (green). This indicates that there are substantial returns to combining policies in terms of individual-level satisfaction.

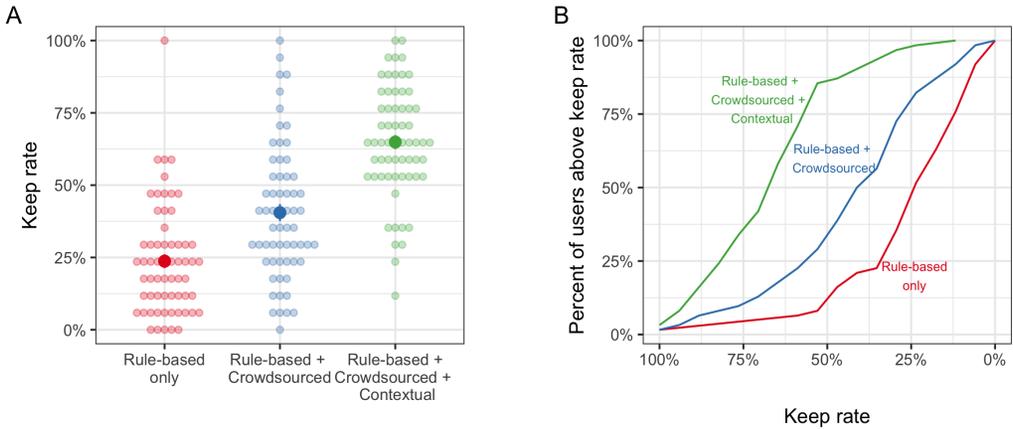

Fig. 8. (A) Keep rates for different policy combinations. Each smaller point represents one participant's average keep rate for perspectives under the indicated combination of policies. The three larger points give the mean for each combination, with error bars showing one standard error. (B) A cumulative version of the same, showing the number of participants at or above a given keep rate under different policy combinations.

*4.2.3 People prefer rule-based perspectives for larger numbers.* We also looked at how the preference for perspectives varies with the magnitude of the focal number. Overall we find that keep rates grow along with the focal number, consistent with responses in the data collection task used to train the contextual model. Interestingly, however, this correlation is not uniform across policies, as depicted in Fig. 9, where we see a statistically significant increase in keep rate by focal number for rule-based perspectives but not for crowdsourced or contextual perspectives. Specifically, we fit a logistic regression for each policy to model if a perspective from that policy was kept as a function of the logarithm of the focal number in each quote. We included fixed effects for the section of the newspaper to try to isolate the effect of the focal number separate from the topic, and random effects at the participant level to account for repeated measures. In each of these models we excluded participants who showed no variation in their reponses (i.e., accepting all or no perspectives from a given policy across all 17 quotes) to avoid convergence issues. Through this analysis we find that there is a sharp rise in the preference for rule-based perspectives as the focal number increases in magnitude ($\hat{\beta}^{per-capita}_{focal} = 0.25$, 95% CI $[0.20, 0.30]$, $z = 10.10$, $p < .001$ ). In contrast, we find that contextualized perspectives are preferred uniformly across focal number magnitudes ($\hat{\beta}^{contextual}_{focal} = 0.00$, 95% CI $[-0.04, 0.03]$, $z = -0.11$, $p = .915$), as are crowdsourced ($\hat{\beta}^{crowdsourced}_{focal} = 0.00$, 95% CI $[-0.04, 0.05]$, $z = 0.18$, $p = .855$), albeit at a lower average keep rate. This rise in preference for rule-based perspectives with the magnitude of the focal number accounts for an overall rise in keep rate with the magnitude of the focal number ($\hat{\beta}^{no\ perspective}_{focal} = -0.21$, 95% CI $[-0.25, -0.17]$, $z = -9.88$, $p < .001$). Further model details are available in Section C.2 of the Appendix.

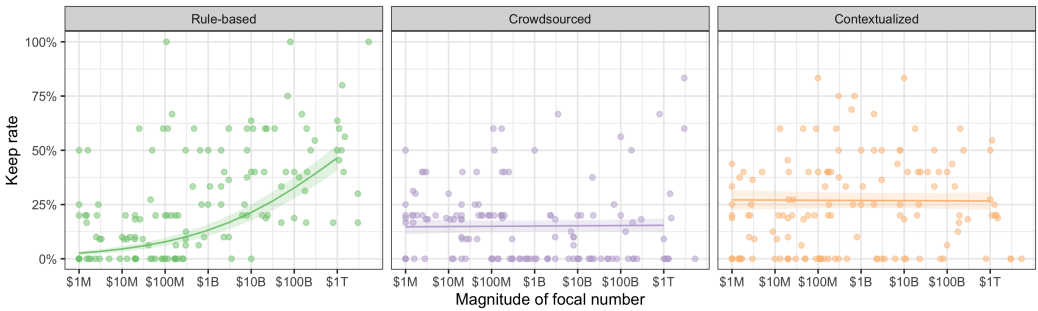

Fig. 9. Keep rate as a function of the magnitude of the focal number in each quote for each policy. Points correspond to quotes, and show the magnitude of the focal number for that quote on the x-axis along with the average keep rate for perspectives under each policy on the y-axis. The lines show marginal means for a mixed effects logistic regression model fit to the points in each panel, with bands showing 95% confidence intervals on the mean.

*4.2.4 Heterogeneity across people and quotes.* Finally, we analyzed whether there is a consistent pattern of choosing perspectives across quotes and participants. We aggregated the keep rate for different types of perspective policies at the quote and participant level, which are shown in Fig. 10 and Fig. 11. As illustrated in the figures, both individuals and quotes show heterogeneity over preferences for perspectives. For instance, there were certain participants who did not find perspectives helpful at all, while other participants may have preference over specific perspective policies (e.g. preference over contextualized perspective). Similarly, people found certain perspective policies more helpful for certain quotes. Our findings open up interesting discussion points on potential personalization of policies based on individuals and quotes, which we discuss in the following section.

## 4.3 Qualitative Follow Up Study

In order to gain more insights into why participants preferred some perspectives over others and how valuable they found automatically generated perspectives overall, we conducted a followup qualitative study. This mirrored the evaluation task described above, but we shortened the number of tasks each worker saw (from 17 to 5, plus the additional attention check) and added a required free text response on each task: "In a sentence or two, please describe the reasons for your selection". At the end of the study we showed participants an animated demonstration of how automatic perspectives could be deployed in a word processor, and asked three final, required questions. First, we asked for a 5 point Likert scale rating about the perceived helpfulness of the system ("If you were routinely in a position of writing articles like these, how helpful would you find a feature like the above?"), along with a text response for explaining their rating. We also asked participants two questions regarding generalizability to other contexts and other measurements ("Besides news articles, where else could a feature like this (that puts numbers into perspective) be useful?" and "Besides dollar amounts, what other kinds of numbers would be useful to put into perspective?").

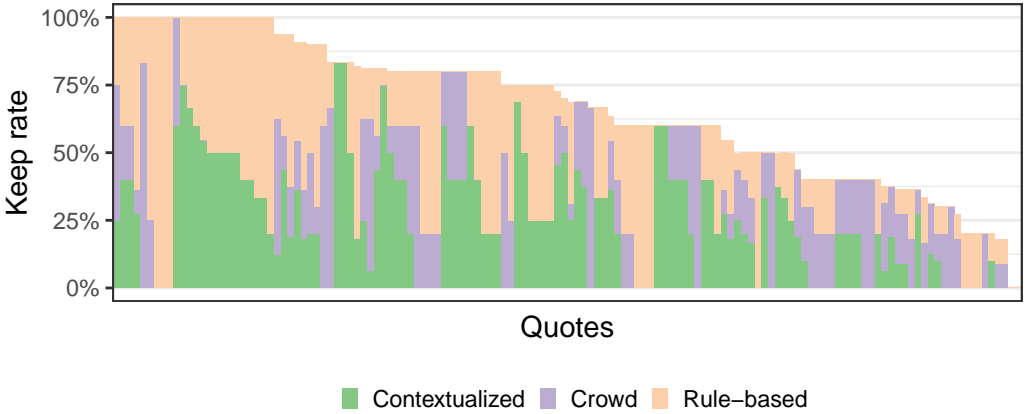

Fig. 10. Keep rate across quotes and policies. Bars represent the 136 individual quotes shown to participants and their keep rate breakdown for the three available policies. Bars are stacked, so that the total height of each column gives the overall fraction of users who selected any perspective for each quote.

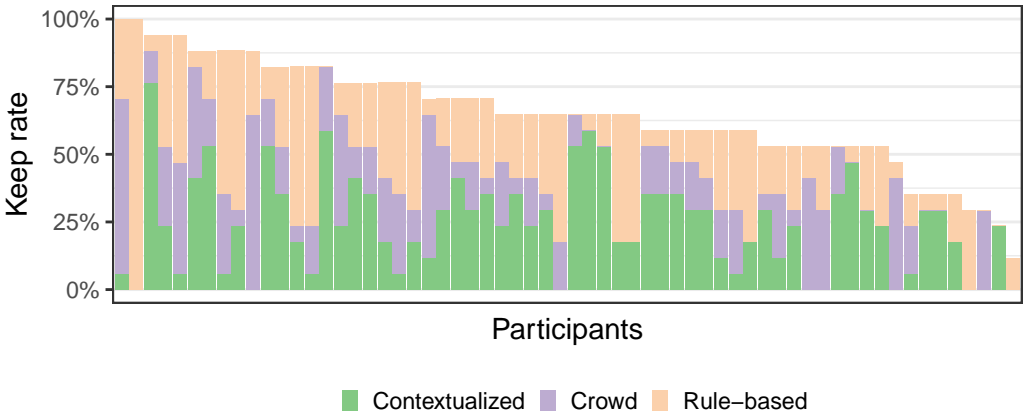

Fig. 11. Keep rate across participants and policies. Bars represent preferences for the 62 participants who passed the attention check. Bars are stacked, so that the total height of each column gives the overall fraction of quotes that each user selected any perspective for.

We recruited 50 U.S. based participants from Amazon Mechanical Turk, with a Masters qualification, 99% approval rate and at least 100 approved HITs. Two participants failed the attention check, and one did not provide meaningful responses to the qualitative questions. We exclude these responses from our analysis.

Overall, the majority of participants (70.8%) rated the system highly (4 or 5 on the 5-point Likert scale), indicating that they would find the feature quite helpful in word processing software.

Participants also reported that the system would be useful for authoring documents beyond news articles, including educational materials, medical reports, business presentations, and financial and legal documents, among others. Beyond dollar amounts, they identified several other types of measurements for which perspectives would be helpful, including weights, distances, temperatures, populations, carbon emissions, and nutritional information.

The free text responses on each of the 5 tasks participants were shown also provided insight into why people prefer some perspectives over others. Thematic analysis conducted on these responses by 3 independent raters ($\kappa_{Fleiss}$ = 0.703) revealed 8 main themes in participants' explanations for choosing particular perspectives: *Contextually Relevant, Familiar Reference, Direct Relation to Self, Good Multiplier, Doubt Correctness, Perspectives not Helpful, Number is Already Clear* and *Other*.

In line with our quantitative results, we found that participants had a mix of reasons for choosing individual perspectives, with a roughly even split between contextual relevance and familiarity. 25% of responses were coded as chosen because they would be helpful with the context (Contextually relevant). Example quotes in this category were "Can give you a better understanding of how much this person makes relative to others in the MLB.", "Because it gives an idea of the size of the company when compared to a larger competitor.". 24% of responses mentioned choosing the item they felt was the most familiar and easily understood (Familiar reference). Some example quotes were "I think this makes it clear how expensive the budget is since most people know how costly private jets are." when choosing to add 'the cost of a high end private jet', and "People generally know that motion pictures cost millions and this is a decent comparison" when choosing to add 'half the budget of a modest motion picture' as a perspective. 15% of responses mentioned preference because it was easier to relate to themselves ("This money comes from taxes, so I think it's easier to understand when shown what the per person cost is.", "$400 per person is a lot. That will help people to understand how large it is.").

While prior work has raised concerns that people are less likely to opt for subtracting information [1, 19], we found in both our first evaluation study (35%) and our follow up study (28%) that participants opted to not add any additional information a considerable amount of the time. Our follow up study revealed different reasons for doing so, with 17% of the responses being due to participants feeling none of the options would be useful, and sometimes even suggesting alternatives (e.g. "All these examples just complicate things.","None of the examples are good, i would go with something like the average cost of HOWEVER MANY homes in the U.S."). 10% of responses were due to participants feeling that the number being considered was already clear and not in need of additional explanation ("There's no need to add any of those random statements, $1.5 million is easy to understand.").

Of note is also the fact that several participants also commented on how the perspectives provided actually helped them learn something new, with some example comments of this type being "This is not only good for perspective and a relevant topic, but it actually enhances the article. It's an interesting piece of information.", "It sounds like an outrageous amount to pay for housing athletes. Seeing this number makes you realize how much we spend on athletes routinely, and makes you understand why people wouldn't think that two billion was wasteful.".

Further details for this study, including screenshots, the full distribution of helpfulness ratings, and a detailed explanation of all coded themes along with the frequencies of different responses can be found in Section D of the Appendix.

# 5 DISCUSSION

## 5.1 Per-Policy Takeaways

In this work, we implemented and compared three policies to automatically generate perspectives for U.S. dollar amounts in various contexts: a rule-based policy, a crowdsourced approach, and a policy for generating contextualized perspectives. As demonstrated by the studies, participants generally found automatically generated perspectives helpful in understanding dollar amounts. Different policies had their own merits and although aggregated keep rates differ, there is no policy that clearly dominates the others. In fact, heterogeneity across both individuals and contexts suggests that there are substantial complementarities between policies, and gains to be had from each.

Rule-based perspectives are very simple to implement, yet they provide a surprisingly strong baseline. On average these perspectives had a keep rate that is competitive with a contextual approach, but which varied with the magnitude of the focal number: users preferred rule-based perspectives for larger focal numbers in the billions or trillions compared to smaller numbers in the millions. This rule-based policy provides a quick and easy way to automatically generate perspectives that users are likely to appreciate and gain some numerical insight from, and which can be adapted to different cultural markets with relative ease.

Likewise, analogs of the rule-based policy for dollar amounts we studied here can be crafted for other types of measurements. For instance, one could express areas in relation to a fixed, well known geographic region (e.g., the United States), calories consumed relative to time spent exercising, or greenhouse gases relative to the carbon emitted on a cross-continental flight.

That said, such policies may suffer in the long term due to a lack of diversity in the perspectives they provide—while we saw no sign of this in our evaluation here, one could imagine that repeatedly seeing the same type of perspective in all contexts could cause people to become insensitive to it or merely desirous of variety.

The crowdsourced system we detailed has benefits in this regard in that it delivers a rather diverse set of manually-vetted reference objects, selected to be both familiar and helpful. That said, in our investigation of quotes in context, the crowdsourced approach led to a lower keep rate compared to the rule-based and contextual approaches. It was also relatively expensive due to the relatively small fraction of suggested references that survive all three of the proposal, verification, and rating phases. Furthermore, this effort and expense scales roughly linearly as the policy is expanded to cover various dimensions. An additional expense is incurred when the crowdsourced database of reference objects needs updating due to the underlying quantities (e.g., a company's market capitalization or the price of various goods) changing over time. Finally, we found that it was more difficult for participants to generate familiar references for extreme focal numbers, which unfortunately are the numbers that people need the most help comprehending.

Contextual perspectives seem to strike a balance between rule-based and crowdsourced perspectives, performing reasonably well across the whole range of focal numbers while offering diverse, context-specific suggestions. Participants in our qualitative study specifically mentioned the helpfulness of adding contextually relevant information in better comprehending points made in quotes. While there is a somewhat significant initial investment required to develop a first contextual policy for one dimension, the marginal effort required to extend this to other dimensions and keep it up to date is comparatively small. For instance, given that we have already parsed Wikidata entries, joined them against Wikipedia traffic data, and used a language model to encode the semantics of reference objects for dollar amounts, we can re-use these same data sources and the existing code to obtain reference objects for other dimensions with minor changes.

## 5.2 Combining Policies

Beyond comparing individual policies directly to one another, this work also demonstrates the value in combining strategies. Specifically, we find a sizable increase in average user satisfaction (as measured by keep rate) by showing perspectives from multiple policies side-by-side. This of course may not always be possible or desirable, for instance in space-constrained settings or situations where one can recommend only one best option to users. In these settings one can imagine adding an additional layer of intelligence to the system in the form of, say, a contextual bandit or personalization service that exploits the heterogeneity across quotes and users to learn which policy to deploy in which settings [17]. For instance, such an approach might learn to use rule-based perspectives for extremely large numbers compared to smaller ones, or that certain users prefer contextual perspectives to rule-based ones. This could also include learning when or when not to display perspectives based on the focal number, as participants seem prefer perspectives for larger focal numbers over smaller ones and do not always find perspectives necessary.

## 5.3 Generalizability

We note that although we explored one type of measurement in this work (dollar amounts), all three policies presented can be readily generalized to other dimensions. Rule-based policies are relatively straightforward, as one can easily define different rules depending on the measurement. Examples of other rule-based policies are presented in Section 2, such as 'gallons per 100 miles' for fuel efficiency [14], and walking miles for calorie consumption [9]. For dollar amounts, although we defined a rule-based policy based on per-capita amounts for the U.S. population, other possible re-expressions could involve other fixed references such as expressing large dollar amounts as fractions of the federal budget. Per-capita rule-based policies can also be adapted to other cultural markets by swapping the reference population from the U.S. to that of other countries.

Our crowdsourced policy is similarly readily generalizable to other measurements, albeit with a somewhat higher incurred cost. We can use our system presented in Section 3.2 as-is, by simply changing the measurement prompts in the generation phase (Section 3.2.1). We have already done so to collect reference objects for area, population, length, height, mass, and volume. Generalizability to other markets is also possible, through country or market specific crowdsourcing participant pools.

As mentioned in Section 5.1, our contextualized policy can also be used for other types of measurements with minimal overhead. We can simply use the existing pipeline for dollar amounts, but apply it to other types of measurements contained in the Wikidata knowledge base, of which there are many. Contextual perspectives can also be adapted to other cultural markets, albeit with some potential challenges. For example, one could imagine adapting the approach presented here to the German population instead of the U.S. by using traffic to the German language edition of Wikipedia as a proxy for familiarity instead of English Wikipedia traffic data. This might work well in that there is a strong correlation between living in Germany and speaking German, however there are other markets for which there is a looser mapping between language and culture (e.g., Spanish is spoken by many different cultures, and countries like Canada or Switzerland are multi-lingual). And while adapting to other dimensions or markets might require collecting additional helpfulness labels to re-train a ranking model, this cost would be substantially lower than sourcing a new set of reference objects from crowd workers.

## 5.4 Use cases and real-world deployment

Regardless of how different policies are selected or combined, high-quality automatically generated perspectives offer many opportunities to improve human-computer interfaces and human-AI collaboration [15]. These range from adding perspectives to reading interfaces (e.g., in web browsers, PDF or e-book readers, etc.), authoring tools (e.g., word processing and other productivity software), social media and journalism platforms (e.g., for fact checking), and in educational settings (e.g., for teaching number sense and scale). The results of this work in particular have informed the design and deployment of a perspectives feature in a widely used online word processor, depicted in Fig. 12. Currently we have implemented both crowdsourced and rule-based perspectives for English-speaking, U.S.-based users, and the overall results that we have seen in the live system are in line with findings presented here. We are currently pursuing a pilot of contextual perspectives pending some engineering work on scalable, real-time embeddings. Future steps include adapting approaches to other cultural markets and incorporating some of the personalization or contextual bandit approaches mentioned above.

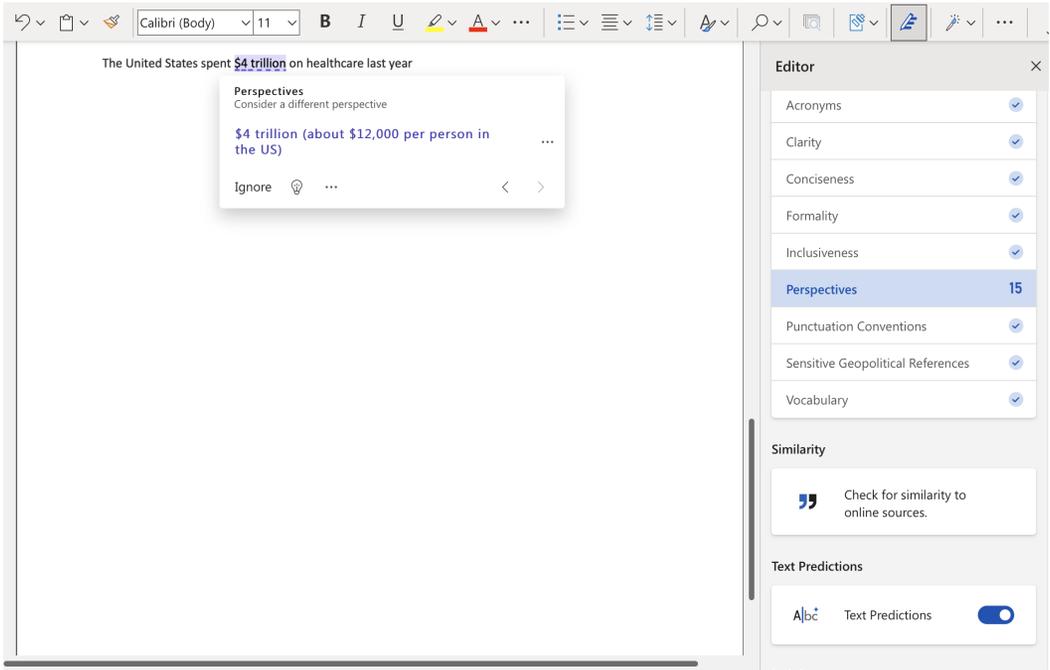

Fig. 12. Perspectives deployed live in a word processor. Similar to spelling or grammar check, when a potentially unfamiliar measurement is detected, a pop-up appears showing a perspective suggestion. Upon clicking the suggestion the perspective is inserted in a parenthetical in the document.

## 5.5 Limitations and future work

Our work of course has several limitations and points to directions for improvements and future research, much of which is guided by our experience from this deployment. First are the challenges, mentioned above, around extending to other dimensions and cultural markets. It is difficult to infer which perspectives will be familiar and helpful in which cultures. In the case of cultures with smaller populations, there can be a lack of data guide these inferences. Second, while the contextual policy we developed works reasonably well, there are several ways in which it could be improved, and so we think of it as providing a lower bound for the capability of such approaches. One simple improvement would be to collect more training data to improve the ranking model, perhaps from a more representative and diverse population of crowd participants. Another would be to use more sophisticated language models to capture semantic similarity, or even to directly generate perspective text, including encoders that are explicitly designed to model numbers for natural language processing [21, 30]. Third is the possibility of exploring other perspective formats beyond multiplicative scaling, such as rank and percentile formats (e.g., "the 3rd highest crime rate among U.S. cities", or "in the top 3% of CEO pay"), or using relative ratios for extreme amounts (e.g., "the ratio between the volume of grain of sand and the Earth"), as discussed in previous research [3, 10].

## 6 CONCLUSION

Overall we hope that this work serves as a demonstration of the benefits of end-to-end systems for automated perspective generation, highlighting the strength of combining different approaches and offering guidelines about the relative costs and benefits of each. Using the rule-based and crowdsourcing methods, we have generated and deployed perspectives into production in a widely used word processor. Going forward, we hope to see more automated processes for generating context-specific perspectives at scale. This seems merited as our data suggest that one way to be inclusive of heterogeneous preferences is to provide readers and writers with a large and diverse set of perspectives.

## A    QUOTE SELECTION FOR CONTEXTUAL PERSPECTIVE MODEL TRAINING AND FINAL EVALUATION

We provide statistics for quotes we selected for training contextual perspective model and for final evaluation, in Fig. 13 and Fig 14 respectively.

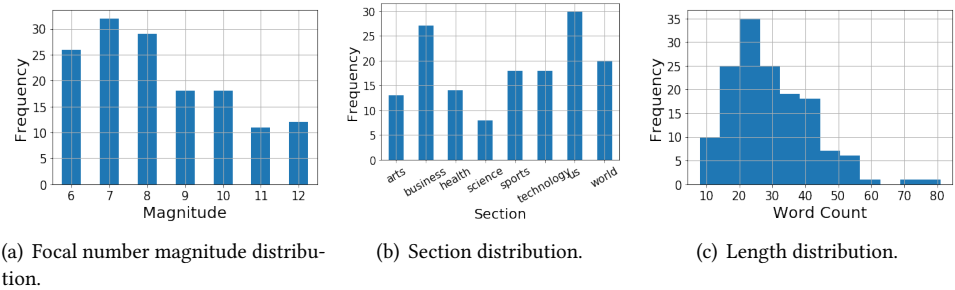

(a) Focal number magnitude distribution.

(b) Section distribution.

(c) Length distribution.

Fig. 13.  Selected quotes for training contextual perspective model.

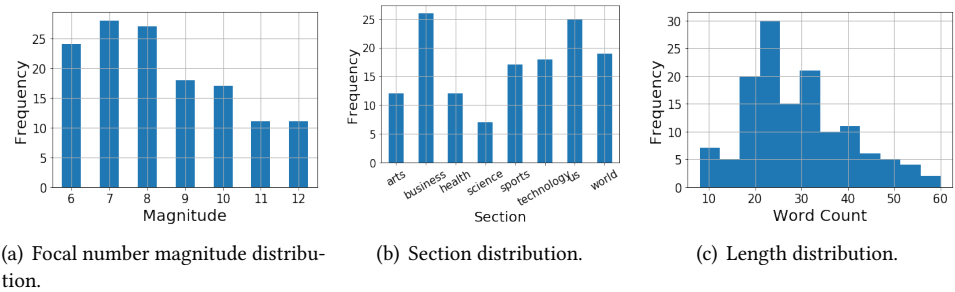

(a) Focal number magnitude distribution.

(b) Section distribution.

(c) Length distribution.

Fig. 14.  Selected quotes for final evaluation.

## B    DETAILS OF CONTEXTUAL PERSPECTIVE MODEL

Each quote has been judged on average by 25 workers. Number of times references belonging to a specific property has been judged in the training data collection procedure is shown in Table 4. The coefficient learnt in contextual perspective model is shown in Fig. 15.

| Property | Number of times judged |
|---|---|
| Dictionary | 11991 |
| Nominal GDP | 486 |
| Nominal GDP per capita | 396 |
| Annual budget | 1141 |
| Cost | 1124 |
| Endowment | 269 |
| Market capitalization | 972 |
| Net profit | 1302 |
| Price | 72 |
| Total assets | 1706 |
| Annual revenue | 23485 |
| Total equity | 796 |

Table 4. Number of times references belonging to a specific property has been judged in the training data collection procedure.

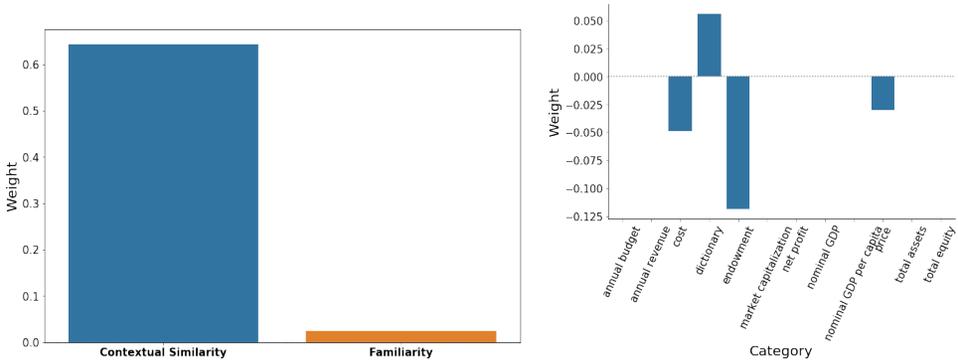

(a) Coefficient learnt for textual relevance and familiarity.

(b) Coefficient learnt for reference objects property type.

Fig. 15. Coefficients for different features in trained contextual perspective model.

| | Model 1: Helpfulness by BERT Score |
|---|---|
| (Intercept) | 0.71*** |
| | (0.05) |
| Bert_score | 1.66*** |
| | (0.09) |
| $R^2$ | 0.15 |
| Adj. $R^2$ | 0.15 |
| Num. obs. | 1800 |

*** $p < 0.001$; ** $p < 0.01$; * $p < 0.05$

Table 5. Linear regression results for helpfulness by BERT score.

## C ADDITIONAL RESULTS EVALUATING POLICIES

## C.1 Combinations of policies on keep rate

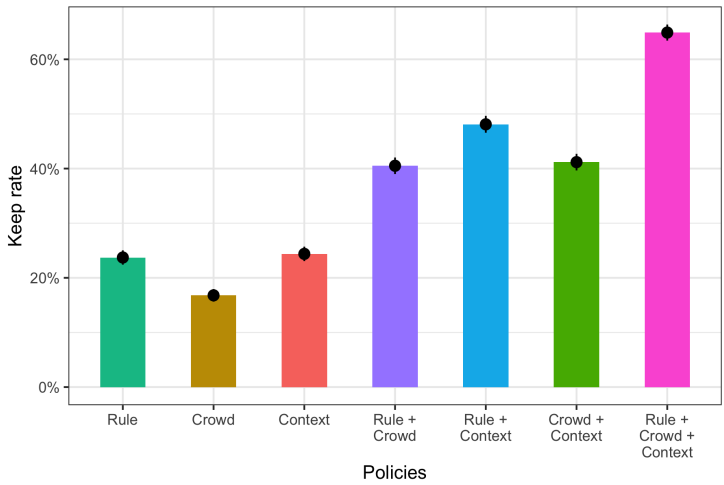

Fig. 16. Average keep rate for all possible combinations of policies. Error bars show one standard error above and below the mean.

## C.2   Logistic regression for keep rate by policy and focal number

|  | Rule-based | Crowdsourced | Contextualized |
|---|---|---|---|
| (Intercept) | $-6.98^{***}$ | $-1.14^{*}$ | $-1.37^{**}$ |
|  | (0.62) | (0.50) | (0.45) |
| log_focal_number | $0.25^{***}$ | 0.00 | $-0.00$ |
|  | (0.03) | (0.02) | (0.02) |
| business | 0.03 | $-0.13$ | $0.68^{*}$ |
|  | (0.41) | (0.35) | (0.33) |
| health | $1.58^{***}$ | $-0.15$ | $-0.10$ |
|  | (0.47) | (0.44) | (0.44) |
| science | 0.13 | $-0.01$ | 0.29 |
|  | (0.51) | (0.44) | (0.43) |
| sports | $-0.05$ | 0.11 | $-0.40$ |
|  | (0.47) | (0.39) | (0.41) |
| technology | 0.48 | $-0.49$ | 0.10 |
|  | (0.43) | (0.39) | (0.36) |
| us | 0.46 | $-0.79^{*}$ | $0.94^{**}$ |
|  | (0.42) | (0.37) | (0.33) |
| world | 0.52 | $-0.39$ | 0.27 |
|  | (0.45) | (0.42) | (0.39) |
| AIC | 887.98 | 838.91 | 1050.10 |
| BIC | 936.57 | 885.75 | 1098.13 |
| Log Likelihood | $-433.99$ | $-409.46$ | $-515.05$ |
| Num. obs. | 952 | 799 | 901 |
| Num. groups: worker_id | 56 | 47 | 53 |
| Var: worker_id (Intercept) | 0.74 | 0.37 | 0.39 |

$^{***}p < 0.001$; $^{**}p < 0.01$; $^{*}p < 0.05$

Table 6. Logistic regression results for keep rate by focal number for each policy, with fixed effects for section.

# D QUALITATIVE STUDY

## Task 1 of 6

Imagine you are writing an article with the following sentence in it and are trying to help your readers best understand the number in the sentence.

*The deal for Qualcomm would be $130 billion including debt.*

Which of the following, if any, would you add to this article to help readers understand the number $130 billion?

| Perspective | Selection |
|---|---|
| about $400 per person in the US | ○ |
| about 6 times the annual revenue of Qualcomm Inc. | ○ |
| about 50% of the value of the gross domestic product (GDP) of South Africa | ● |
| I would not add any of these perspectives to the article. | ○ |

In a sentence or two, please describe the reasons for your selection *

[                                                                        ]

Next

Fig. 17. Screenshot of the qualitative study task. The design is similar to the first evaluation study, with the addition of a question that appears after participants select an option.

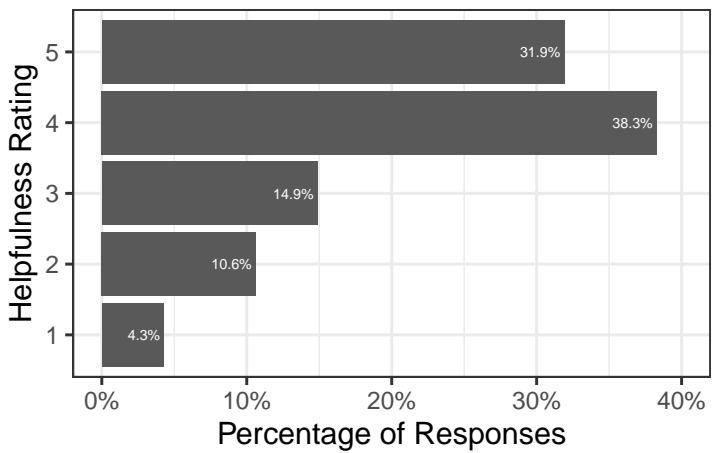

Fig. 19. Distribution of ratings for the 5 point Likert scale question regarding the helpfulness of the perspectives system for writing articles. 1 indicates *Not Helpful*, 5 indicates *Very Helpful*. The majority of participants (70.8%) rated the system highly (4 or 5).

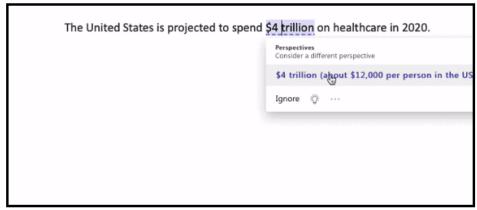

**Thanks for participating!**

We have a few final questions for you. You must answer these questions and click submit below to complete this HIT and get paid.

The animation below shows how a feature that suggests perspectives for unfamiliar numbers might work. When writing articles, it underlines numbers and allows you to insert perspectives in your document.

If you were routinely in a position of writing articles like these, how helpful would you find a feature like the above? *

Not Helpful  ○ 1  ○ 2  ○ 3  ○ 4  ○ 5   Very Helpful

In a few sentences, please tell us more explaining your ratings to the question above: *

Besides news articles, where else could a feature like this (that puts numbers into perspective) be useful?*

Besides dollar amounts, what other kinds of numbers would be useful to put into perspective?*

Please enter any additional feedback below.

You must click this button to complete the HIT

Fig. 18. Screenshot of the survey questions at the end of the qualitative study task. We provided participants with an animated demonstration of how the system works on a word processor. We asked A 5-point likert scale question for rating helpfulness and two open ended questions about generalizability to other measurements and contexts.

| Themes | Description | Chose Perspective | |
|---|---|---|---|
| | | *Yes* | *No* |
| Contextually relevant | Helps with context of text. | 25% | 0% |
| Familiar reference | Easily understood and visualized, familiar to most people. | 24% | 0% |
| Direct relation to self | Mentions personal benefit from the comparison. | 15% | 0% |
| Good multiplier | Multiplier used is helpful, compared to other options. | 1% | 0% |
| Perspectives not helpful | None of the options are useful. | 0% | 17% |
| Number is already clear | Number doesn't need to be put in perspective. | 0% | 10% |
| Doubt correctness | Doubts correctness of some (or all) options. | 2% | 1% |
| Other | Other reasons. | 5% | 0% |
| | **Overall:** | 72% | 28% |

Table 7. Themes identified through thematic analysis of text responses in the qualitative study, using 3 independent raters, along with overall distributions.